\documentclass[preprint,showpacs,showkeys,preprintnumbers,amsmath,amssymb]{revtex4}

\usepackage{graphicx}
\usepackage{dcolumn}
\usepackage{bm}

\begin{document}


\title{Superintegrability of the Caged Anisotropic Oscillator}

\author{N. W. Evans}%
\email{nwe@ast.cam.ac.uk}
\author{P. E. Verrier}
\email{pverrier@ast.cam.ac.uk}
\affiliation{%
Institute of Astronomy, Madingley Rd, University of Cambridge, CB3 0HA, UK
}%

\date{\today}


\begin{abstract}
  We study {\it the Caged Anisotropic Harmonic Oscillator}, which
  is a new example of a superintegrable, or accidentally degenerate
  Hamiltonian. The potential is that of the harmonic oscillator with
  rational frequency ratio ($l:m:n$), but additionally with barrier
  terms describing repulsive forces from the principal planes. This
  confines the classical motion to a sector bounded by the principal
  planes, or cage. In 3 degrees, there are five isolating integrals of
  motion, ensuring that all bound trajectories are closed and strictly
  periodic. Three of the integrals are quadratic in the momenta, the
  remaining two are polynomials of order $2(l+m-1)$ and $2(l+n-1)$ .
  In the quantum problem, the eigenstates are multiply degenerate,
  exhibiting $l^2m^2n^2$ copies of the fundamental pattern of the
  symmetry group $SU(3)$.
\end{abstract}

\keywords{Classical mechanics, quantum theory, integration}

\maketitle


\section{\label{sec:intro}Introduction}

The subject of accidental degeneracy has fascinated physicists since
the early days of the quantum theory. It has long been known that the
trajectory in an $r$-squared force (the Coulomb or Kepler problem) is
a conic section, so that every bound orbit is an ellipse and therefore
closed and strictly periodic. The advent of the quantum theory saw the
deduction of the eigenfunctions for the Hydrogen atom and thence the
realization that the bound states of the Coulomb problem are
degenerate ~\cite{Pa26,Fo35, Ba36}. This accidental degeneracy is a
consequence of a hidden symmetry group SO(4) that is not manifest to
the eye. It causes all bound trajectories to be closed in the
classical problem.

There is a fundamental connection between accidental degeneracy and
the separability of the Schr\"odinger or the Hamilton-Jacobi equations
in more than one coordinate system and therefore the existence of
additional conserved quantities or integrals of motion. This is
mentioned in a number of the famous texts of the old quantum theory --
such as Born's {\it Mechanics of the Atom} and Sommerfeld's {\it
  Atomic Structure and Spectral Lines}.  For example, the Coulomb
problem is separable in both spherical polar and rotational parabolic
coordinates. The former leads to the conservation of the angular
momentum vector, the latter to the conservation of the
Laplace-Runge-Lenz vector. The quantum mechanical operators close to
form the algebra of SO(4) (see, for example, \cite{Ar76} for a
review). The accidental degeneracy is a consequence of additional
integrals of motion \cite{We93} and so such systems are often called
superintegrable \cite{Te05}.

Systematic investigations of all the possible combinations of
coordinate systems for which the Schr\"odinger and Hamilton-Jacobi
equation can separate have now been carried out~\cite{Fr65, Ma67,
  Ev90}. In three degrees of freedom, this yields 13 distinct
superintegrable systems, all of which have classical integrals of
motion or quantum operators that are quadratic in the canonical
momenta and all of which exhibit accidental degeneracy. This includes
familiar systems such as the Coulomb problem and the isotropic
harmonic oscillator.

Nonetheless, this does not provide a comprehensive explanation of the
phenomenon of superintegrability. For example, in three degrees of
freedom, the Hamiltonian of the anisotropic harmonic oscillator with
rational frequency ratio is
\begin{equation}
H = \frac{1}{2}(p_x^2 + p_y^2 + p_z^2) + k(l^2 x^2 + m^2 y^2 + n^2
z^2), 
\label{eqn:aniso}
\end{equation}
where $l, m$ and $n$ are integers and $k$ is a constant. The
Hamilton-Jacobi or Schr\"odinger equations clearly separate in
rectangular Cartesians. If $l:m :n = 2:1:1$, then the Hamiltonian also
separates in the rotational parabolic and elliptic cylindrical
coordinate systems~\cite{Ev90}, giving rise to additional conserved
quantities and accidental degeneracy. However, if $l + m +n > 4$, then
the Hamiltonian is still superintegrable~\cite{Ja40,AW02,BP96}, even
though it now only separates in rectangular Cartesians. Further
examples of systems which are superintegrable but not separable in
more than one coordinate system include the Calogero-Moser
problem~\cite{Ad77, Wo83} and the generalized Coulomb
problem~\cite{VE08}.

The purpose of this paper is to introduce another superintegrable
Hamiltonian, closely related to (\ref{eqn:aniso}), namely
\begin{equation}
H = \frac{1}{2}(p_x^2 + p_y^2 + p_z^2) + k(l^2 x^2 + m^2 y^2 + n^2
z^2) + \frac{k_1}{x^2} + \frac{k_2}{y^2} + \frac{k_3}{z^2}
\label{eqn:h}
\end{equation}
We shall refer to this as ``the Caged Anisotropic Oscillator'', as the
presence of the barrier terms confines the motion to an octant defined
by the principal planes. When $l=m=n=1$, this becomes the
Smorodinsky-Winternitz system, on which there is an extensive
literature ~\cite{Fr65,Ev91,Ba07}.

Like the Smorodinsky-Winternitz system, the Caged Anisotropic
Oscillator has five integrals of motion and it exhibits accidental
degeneracy in quantum mechanics. However, unlike the
Smorodinsky-Winternitz system, the Hamilton-Jacobi and Schr\"odinger
equations only separate in rectangular Cartesians.  In this paper, we
demonstrate that the Hamiltonian (\ref{eqn:h}) is superintegrable
using the method of projection in \S 2. Then we discuss both the
classical and quantum problems in some detail in \S 3 and \S 4
respectively.


\section{Proof of Superintegrability}
\label{sec:proof}

Let us start with the observation that the commensurate anisotropic
oscillator in $N$ dimensions possesses $2N-1$ functionally independent
integrals of motion, equal to the $N$ energies of each individual
oscillator and the $N-1$ phases differences between them
\cite{AW02}. In six dimensions, we have
\begin{equation}
\label{eqn:6d}
H_6 = \sum_{i=0}^6 \left( \frac{1}{2} p_i^2 + k n_i^2 s_i^2 \right)
\end{equation}
where the $n_i$ are positive integers and the $s_i$ are Cartesian
coordinates. If we now introduce coordinates $(x,y,z,\theta_x,
\theta_y, \theta_z)$ according to
\begin{eqnarray}
s_1 &=& x \cos\theta_x, \qquad s_2 = x \sin \theta_x \nonumber\\
s_3 &=& y \cos\theta_y, \qquad s_4 = y \sin \theta_y \nonumber\\
s_5 &=& z \cos\theta_z, \qquad s_6 = z \sin \theta_z \nonumber
\label{eqn:transf}
\end{eqnarray}
and let the frequencies $n_1 = n_2 = l$, $n_3 = n_4 = m$, $n_5=n_6=n$,
we have the Hamiltonian
\begin{equation}
H = \frac{1}{2}\left( p_x^2 + p_y^2 + p_z^2 + \frac{p_{\theta_x}^2}{x^2}
+ \frac{p_{\theta_y}^2}{y^2} + \frac{p_{\theta_z}^2}{z^2}\right) + k(l^2 x^2 + m^2 y^2 + n^2 z^2)
\end{equation}
Since the coordinates $(\theta_x, \theta_y, \theta_z)$ are ignorable
we can set the conjugate momenta to constants. Making the
substitutions $p_{\theta_x}^2 = k_1$, $p_{\theta_y}^2 = k_2$,
$p_{\theta_z}^2 = k_3$ gives us the Caged Harmonic Oscillator
Hamiltonian (\ref{eqn:h}).  

For the Hamiltonian in 6 dimensions given by~(\ref{eqn:6d}), every
bound trajectory is closed. Similarly, in the reduced 3 degrees of
freedom Hamiltonian~(\ref{eqn:h}), every bound trajectory is also
closed and the system is still superintegrable.

\begin{figure}[ht]
\centering
\includegraphics[width=\textwidth]{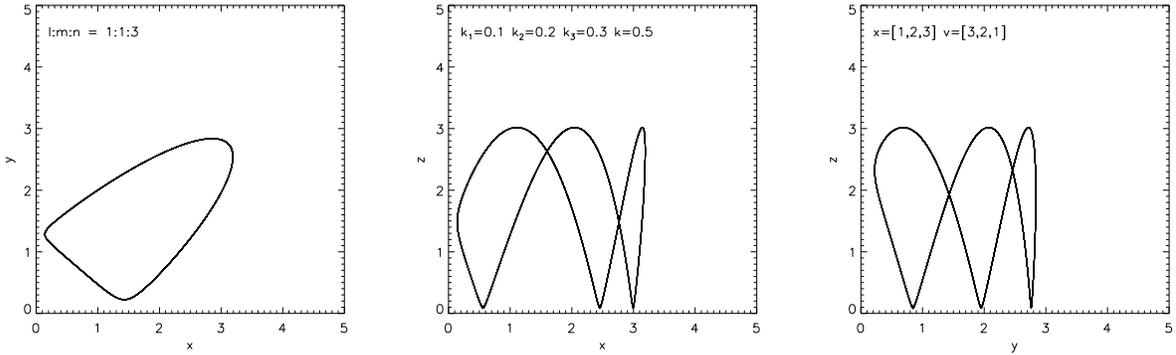}
\caption{A typical orbit for a frequency ratio of $l:m:n = 1:1:3$.}
\label{fig:orbit3}
\end{figure}
\begin{figure}[ht]
\centering
\includegraphics[width=\textwidth]{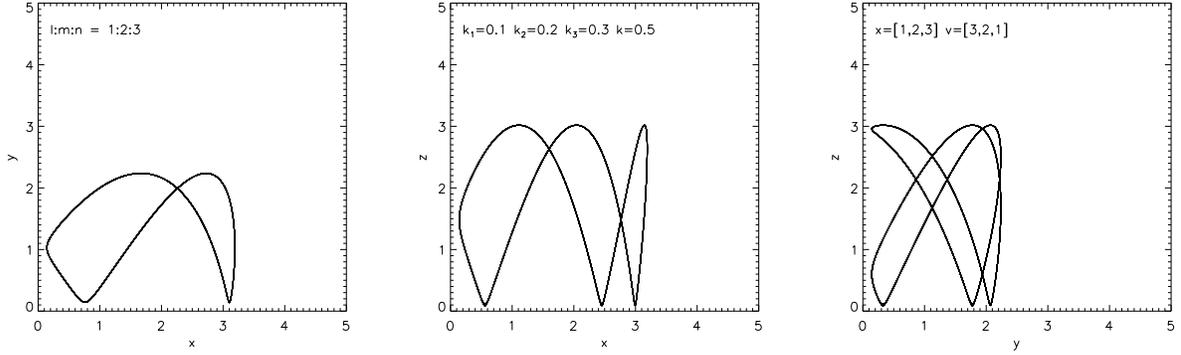}
\caption{A typical orbit for a frequency ratio of $l:m:n = 1:2:3$.}
\label{fig:orbit4}
\end{figure}

\section{Classical Mechanics}
\label{sec:int}

\subsection{The Integrals of Motion}

The Hamiltonian~(\ref{eqn:h}) clearly separates in rectangular
Cartesian coordinates to give the first three integrals of motion as
the three energies of oscillation
\begin{eqnarray}
I_1 &=& \frac{1}{2} p_x^2 + k l^2 x^2 + \frac{k_1}{x^2}\\
I_2 &=& \frac{1}{2} p_y^2 + k m^2 y^2 + \frac{k_2}{y^2}\\
I_3 &=& \frac{1}{2} p_z^2 + k n^2 z^2 + \frac{k_3}{z^2}
\end{eqnarray}
As the system is separable in these coordinates, it is easy to see
that the trajectories of the orbits are (c.f. \cite{Ev90})
\begin{eqnarray}
x^2 &=& \frac{I_1}{2 l^2 k} + \left( \frac{I_1^2}{4 l^4 k^2} - \frac{k_1}{l^2 k} \right)^{1/2} \cos (\sqrt{8 k} l (t-t_0) ) \\
y^2 &=& \frac{I_2}{2 m^2 k} + \left( \frac{I_2^2}{4 m^4 k^2} - \frac{k_2}{m^2 k} \right)^{1/2} \cos (\sqrt{8 k} m (t-t_0) + c_1) \\
z^2 &=& \frac{I_3}{2 n^2 k} + \left( \frac{I_3^2}{4 n^4 k^2} - \frac{k_3}{n^2 k} \right)^{1/2} \cos (\sqrt{8 k} n (t-t_0) + c_2) 
\end{eqnarray}
where $t_0$ and the $c_i$ are constants. The remaining two integrals
are the phase differences between the orbits, say $c_1$ and $c_2$. If
we say too that $|m-l| < |n-l| < |n-m|$, then the integrals are also
of lowest order possible in the momenta. First, let us define
\begin{eqnarray}
\xi  \doteq \frac{x^2-\alpha}{A} &=&  \cos (\sqrt{8 k} l (t-t_0) ) \\
\eta \doteq \frac{y^2-\beta}{B} &=&  \cos (\sqrt{8 k} m (t-t_0) + c_1) \\
\zeta \doteq \frac{z^2-\gamma}{C} &=&  \cos (\sqrt{8 k} n (t-t_0) + c_2) 
\end{eqnarray}
where $\alpha = I_1/(2 l^2 k)$, $\beta = I_2/(2 m^2 k)$ and $\gamma =
I_3/(2 n^2 k)$ and
\begin{eqnarray}
A  & = & \left( \frac{I_1^2}{4 l^4 k^2} - \frac{k_1}{l^2 k}
\right)^{1/2} , \qquad\qquad B = \left( \frac{I_2^2}{4 m^4 k^2} -
\frac{k_3}{m^2 k} \right)^{1/2} , \nonumber \\
C & = & \left( \frac{I_3^2}{4 n^4 k^2} - \frac{k_3}{n^2 k} \right)^{1/2}
\end{eqnarray}
The derivation of both integrals is similar and will be demonstrated
with the case of $c_1$.  The first phase difference is given by (c.f.,
the discussion of the anisotropic oscillator in \cite{BP96})
\begin{equation}
c_1 = \arccos \eta - \frac{m}{l} \arccos \xi
\end{equation}
Taking the cosine gives
\begin{eqnarray}
\cos (l c_1) &=& \cos (l \arccos\eta) \cos ( m \arccos \xi ) + \sin (l \arccos \eta) \sin (m \arccos \xi) \nonumber \\
		   &=& T_l(\eta) T_m(\xi) + \frac{\dot{\xi}\dot{\eta}}{8 k m^2 l^2} T'_l(\eta) T'_m(\xi)
\end{eqnarray}
where $T_l$ and $T_m$ are the Chebyshev polynomials of the first kind
~\cite{Er} and $T'_l$ and $T'_m$ are their derivatives with respect to
the arguments $\eta$ and $\xi$. The time derivatives $\dot{\xi}$ and
$\dot{\eta}$ are equal to $2 x p_x/A$ and $2 y p_y / B$ respectively.
It is more convenient to express the integral as
\begin{equation}
I_4 = (2k)^{l+m} A^m B^l \cos (l c_1)
\label{eqn:integralone}
\end{equation}
which is of order $2(l+m)$ in the momenta, but can be reduced
to order $2(l+m-1)$ since the two highest powers of the momenta can be
removed though a combination of the energy integrals.

The corresponding integral for the second phase difference $c_2$ is
\begin{equation}
I_5 = (2k)^{l+n} A^n C^l \cos (l c_2)
\label{eqn:integraltwo}
\end{equation}
where
\begin{equation}
\cos (l c_2) = T_l(\zeta) T_n(\xi) + \frac{\dot{\xi}\dot{\zeta}}{8 k n^2 l^2} T'_l(\zeta) T'_n(\xi)
\end{equation}
which can be reduced to order $2(l+n-1)$ in the momenta. It is easy to
verify that both $I_4$ and $I_5$ are integrals of motion by showing
that the Poisson bracket with the Hamiltonian vanishes. They are also
functionally independent, as may be verified by computing the rank of
the appropriate Jacobian.

\begin{figure}[ht]
\centering
\includegraphics[width=\textwidth]{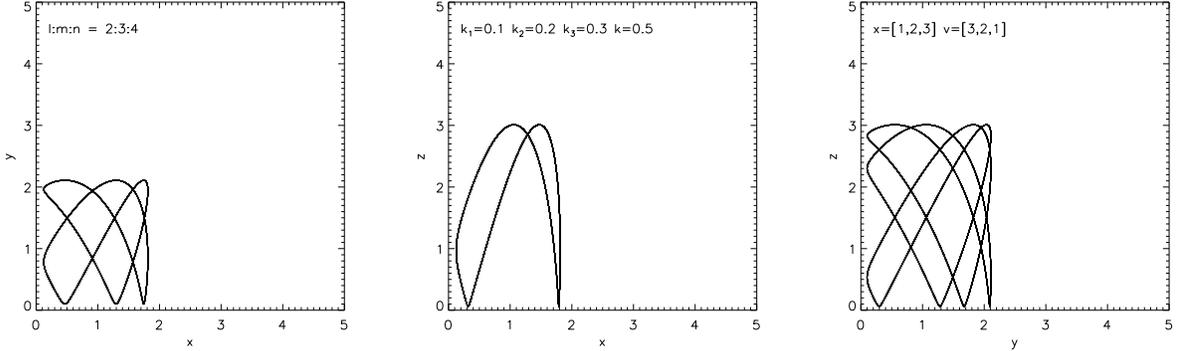}
\caption{A typical orbit for a frequency ratio of $2:3:4$.}
\label{fig:orbit5}
\end{figure}
\begin{figure}[ht]
\centering
\includegraphics[width=\textwidth]{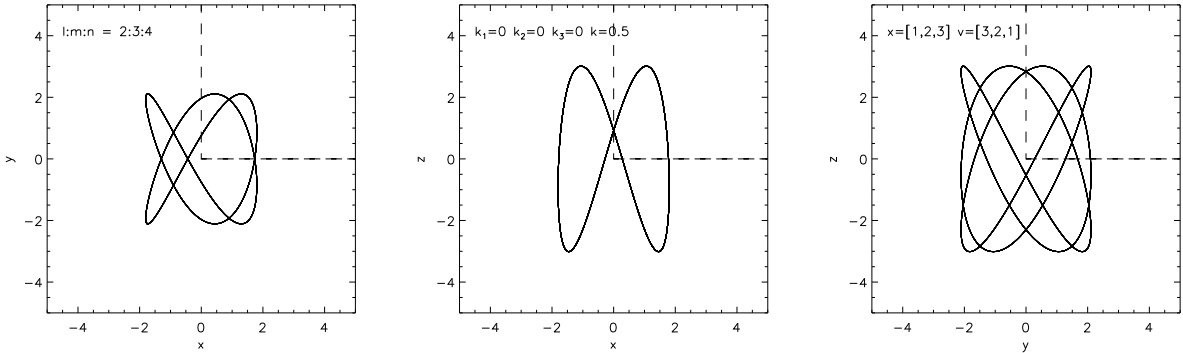}
\caption{A typical orbit for a frequency ratio of $2:3:4$ with no
  potential barriers ($k_1= k_2 = k_3 =0$). The dotted lines show the
  field of view of the corresponding orbit with the potential barriers
  as shown in Fig \ref{fig:orbit5}.}
\label{fig:orbit7}
\end{figure}


\subsection{The Group Theoretic Approach}

Rodriguez et al. \cite{Ro08} have also recently examined this system,
and derived the classical integrals. Ingeniously, they look for
invariants under $SO(2) \times SO(2) \times SO(2)$ corresponding to
the transformation~(\ref{eqn:transf}). They then search for
combinations of these invariants that commute with the
Hamiltonian~(\ref{eqn:6d}) under the Poisson bracket. Such quantities
will also necessarily be integrals of the reduced Hamiltonian of the
Caged Anisotropic Oscillator. Rodriguez et al find integrals of motion
that are rational functions in the momenta, but here we show how to
adapt their method to give integrals that are polynomial in the
momenta.

Let us start by introducing the complex variables
\begin{eqnarray}
z_1 &=& p_1 - i\ell\sqrt{2k} s_1, \qquad z_2 = p_2 -i \ell \sqrt{2k} s_2,\nonumber\\
z_3 &=& p_3 - im\sqrt{2k} s_3, \qquad z_4 = p_4 -i m\sqrt{2k} s_4,\nonumber\\
z_5 &=& p_5 - in\sqrt{2k} s_5, \qquad z_6 = p_6 -i n\sqrt{2k} s_6,\
\end{eqnarray}
so that the Hamiltonian~(\ref{eqn:6d}) is just
\begin{equation}
H = \frac{1}{2} \sum_{i=1}^6 | z_i |^2.
\end{equation}
Now, following Rodriguez et al, we look for invariants under the
generators of rotations in the $(s_1,s_2)$, $(s_3, s_4$) and
($s_5,s_6$) planes. For the ($s_1,s_2$) plane, they include
\begin{eqnarray}
z_1^2 + z_2^2,\qquad {\bar z_1^2} + {\bar z_2^2},\qquad |z_1|^2 + |z_2|^2 .
\end{eqnarray}
with similar results holding for the ($s_3,s_4$) and ($s_5,s_6$)
planes.  Expressions like $|z_1|^2 + |z_2|^2$ clearly commute with the
Hamiltonian~(\ref{eqn:6d}) and are just the separable energies in the
oscillation in the coordinate directions. The remaining quantities do
not commute with~(\ref{eqn:6d}), but it is possible look for an
invariant that is a function of the two expressions ${\bar z_1}^2 +
{\bar z_2}^2$ and $z_3^2 + z_4^2$ that does. Therefore, we require
that
\begin{equation}
\{ H_6, f({\bar z_1}^2 + {\bar z_2}^2, {z_3}^2 + {z_4}^2\} =0.
\end{equation}
Inserting $H_5$ from~(\ref{eqn:6d}), this gives the complex invariant
\begin{equation}
R = ({\bar z_1}^2 + {\bar z_2}^2)^m(z_3^2 +z_4^2)^\ell
\end{equation}
whose real part
\begin{equation}
R + {\bar R} =  ({\bar z_1}^2 + {\bar z_2}^2)^m(z_3^2 +z_4^2)^\ell +
 (z_1^2 + z_2^2)^m({\bar z_3}^2 + {\bar z_4}^2)^\ell
\end{equation}
is a polynomial of order $2(\ell +m)$.  Modulo an unimportant overall
numerical factor, it is the same as the polynomial invariant found
earlier in eq~(\ref{eqn:integralone}).  Similarly, the
invariant~(\ref{eqn:integraltwo}) is just
\begin{equation}
 ({\bar z_1}^2 + {\bar z_2}^2)^n(z_5^2 +z_6^2)^\ell +
 (z_1^2 + z_2^2)^n({\bar z_5}^2 + {\bar z_6}^2)^\ell
\end{equation}
up to a numerical factor.

\begin{figure}[ht]
\centering
\includegraphics[width=\textwidth]{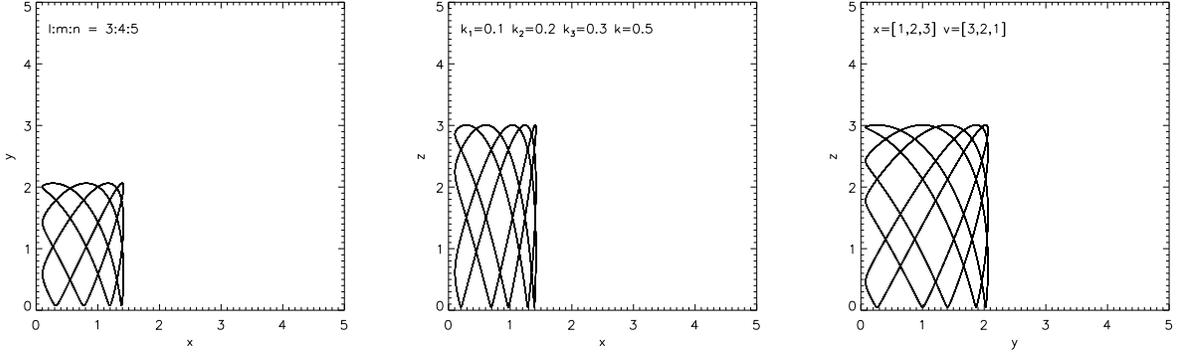}
\caption{A typical orbit for a frequency ratio of $3:4:5$.}
\label{fig:orbit6}
\end{figure}

\begin{figure}[ht]
\centering
\includegraphics[width=\textwidth]{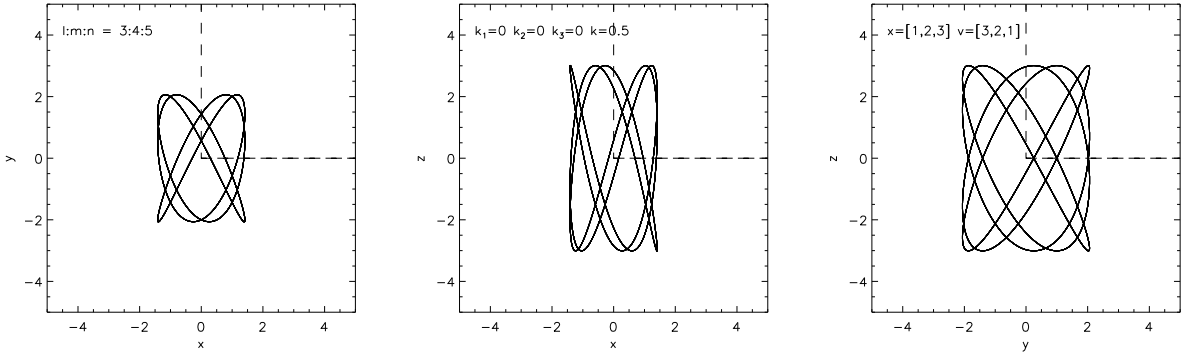}
\caption{A typical orbit for a frequency ratio of $3:4:5$ with no
  potential barriers ($k_1 = k_2 = k_3 =0$).  The dotted lines show
  the field of view of the corresponding orbit with the potential
  barriers as shown in Fig \ref{fig:orbit6}.}
\label{fig:orbit8}
\end{figure}


\subsection{The Orbits}
\label{sec:numexpt}

It is interesting to plot out the orbit of a particle in the
potential. As the Hamiltonian is superintegrable, all bound orbits
must be closed curves. Using a standard Bulirsch-Stoer integrator
\cite{NR} to solve the equations of motion, some example orbits are
plotted. These are shown for various frequency ratios in Figs
\ref{fig:orbit3}, \ref{fig:orbit4}, \ref{fig:orbit5} and
\ref{fig:orbit6}. The orbits are confined to a box, defined by the
limits $\alpha \pm A$, and similar. Figs \ref{fig:orbit7} and
\ref{fig:orbit8} show the corresponding cases to Figs \ref{fig:orbit5}
and \ref{fig:orbit6}, but with no potential barriers. Here, the
trajectories are the well-known Lissajous figures~\cite{Sy60}, and it
can be seen how the orbit in the general case is a reflection and
slight distortion in the $x$, $y$ and $z$ axes.


\section{Quantum Mechanics}

\begin{figure}[ht]
\centering
\includegraphics[width=\textwidth]{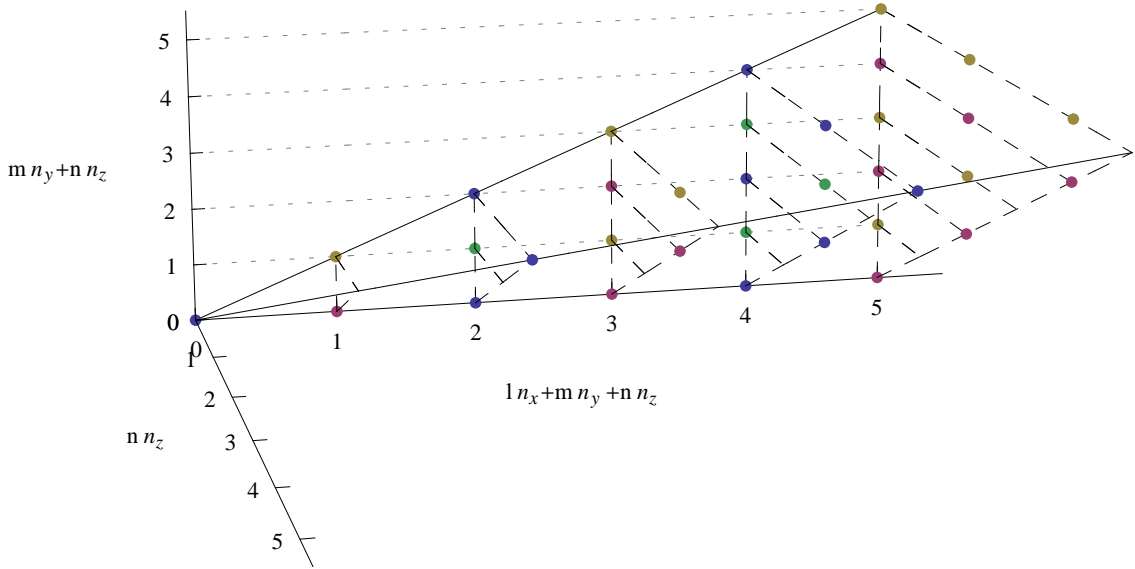}
\caption{Energy levels for the frequency ratio $1:1:2$. Objects belong
  to different sets of ($r_1,r_2,r_3$) values are shown in different
  color. For each color, the degeneracies are the dimensions of the
  irreducible representations of SU(3).}
\label{fig:en2}
\end{figure}

The Schr\"odinger equation is separable in rectangular Cartesians, and
reads:
\begin{equation}
\left(-\nabla^2 + \sum_{i=1}^3 \Big[2 \omega_i^2 kx_i^2 + \frac{2k_i}{x_i^2}\Big]\right)\Psi=2E\Psi
\end{equation}
where $\omega_i$ are the integer multipliers of the frequencies,
corresponding to $l,m,n$ in the previous section, and $\hbar=1$. The
separable solution has wavefunction $\Psi=\prod_{i=1}^3\psi_{n_i}$
where the individual wavefuntions are (c.f., \cite{Fr65,Wi68,Ev91})
\begin{equation}
\psi_{n_i}(x_i) = N_{n_i} e^{-(w_i  \sqrt{k/2} ) x^2_i} x_i^{1/2 \pm \nu_i} L_{n_i} ^{\pm \nu_i}( w_i \sqrt{2k} x^2_i)
\end{equation}
where $N_{n_i}$ is the normalisation constant given by
\begin{equation}
N_{n_i} = w_i^{1/2} (2k)^{1/4} \sqrt{ (2 w_i^2 k)^{\pm \nu_i/2} \Gamma(n_i +1) / \Gamma(n_i +1 \pm \nu_i) }
\end{equation}
and $L_{n_i}^\nu$ are associated Laguerre polynomials, the $\Gamma$
are Gamma functions and $\nu_i = \frac{1}{2} (1 + 8 k_i)^{1/2}$. The
quantised energy is given by
\begin{equation}
E =  2 \sqrt{2k} \sum_{i=1}^{3}   w_i \left( n_i + \frac{1}{2} \pm
  \frac{\nu_i}{2} \right)
\label{eq:qenergy}
\end{equation}
The degeneracy of each energy level with quantum number $N=w_1 n_1 +
w_2 n_2 + w_3 n_3 $ is therefore the same as that of the three
dimensional anisotropic harmonic oscillator with rational frequency
ratio $w_1:w_2:w_3$ (listed for example in \cite{Bo97})
In the simplest case, if the frequency ratio is $1:1:n$ then the
degeneracy is given by
\begin{equation}
g(N) = \Bigg( \left \lbrack\frac{N}{n} \right \rbrack +1 \Bigg) \Bigg(N + 1 - \frac{n}{2} \left\lbrack \frac{N}{n} \right\rbrack \Bigg)
\end{equation}
where $\lbrack N/n \rbrack$ denotes the integer part of $N/n$. The
allowed states for three degrees of freedom and the frequency ratio
$1:1:2$ are shown in Fig.~\ref{fig:en2}.

To look at the group structure, the annihilation and creation
operators can be constructed as
\begin{eqnarray}
b_i &=& \frac{-1}{4 w_i \sqrt{2k}} \left( 2 w_i \sqrt{2k} x_i \frac{\partial}{\partial x_i} + 2 w_i^2 k x_i^2 -\frac{2 k_i}{x^2_i} + w_i \sqrt{2k} + \frac{\partial^2}{\partial x_i^2} \right) \\
b_i^\dagger & = & \frac{-1}{4 w_i \sqrt{2k}} \left( - 2 w_i \sqrt{2k} x_i \frac{\partial}{\partial x_i} + 2 w_i^2 k x_i^2 -\frac{2 k_i}{x^2_i} - w_i \sqrt{2k} + \frac{\partial^2}{\partial x_i^2} \right) 
\end{eqnarray}
which annihilate and create quanta of energy in the $i$ direction,
that is
\begin{eqnarray}
\lbrack H , b_i \rbrack &=& - 2 w_i \sqrt{2k} b_i \\
\lbrack H , b_i^\dagger \rbrack &=& 2 w_i \sqrt{2k} b_i^\dagger 
\end{eqnarray}
and, representing $\psi_{n_i}$ as $\vert n_i \rangle$, act in the
following way
\begin{eqnarray}
b_i \vert n_i \rangle &=& \sqrt{n_i (n_i \pm \nu_i)} \vert n_i -1 \rangle \\
b_i^\dagger \vert n_i \rangle &=& \sqrt{(n_i+1) (n_i \pm \nu_i+1)} \vert n_i +1 \rangle 
\end{eqnarray}
which are identical to those for the Smorodinsky-Winternitz system
given by \cite{Ev91}. As such the number operator given by $\hat{n_i}
= \frac{1}{2}(\lbrack b_i,b^\dagger_i \rbrack \mp \nu_i -1)$ can be
used to again to construct the operators
\begin{equation}
T_{ij} = \frac{1}{2}\lbrace b_i^\dagger (\hat{n_i}\pm\nu_i+1)^{-1/2},b_j(\hat{n_j}\pm\nu_j)^{-1/2} \rbrace
\end{equation}
which close under commutation
\begin{equation}
\lbrack T_{ij}, T_{rs} \rbrack = \delta_{jr} T_{is} - \delta_{is} T_{rj}
\end{equation}
and give the Lie algebra u(3).

In the case of the isotropic harmonic oscillator, it is well known
that the degeneracy of the $N$th energy level is
$(N\!+\!1)(N\!+\!2)/2$, which corresponds to the dimensions of the
irreducible representations of $SU(3)$ . Even though the symmetry
group of the anisotropic harmonic oscillator is also $SU(3)$, it is no
now longer the case that the degeneracy levels follow the pattern $1,
3, 6, 10, 15 ...$. This was already noted as a complication by Jauch
\& Hill~\cite{Ja40}, and there have been a number of possible
resolutions proposed in the literature~\cite{De63,Lo73,Ki73,Ro89,Bo97}

Following the lines of argument put forward in~\cite{Lo73}, we
define
\begin{equation}
{\tilde n}_1 =  n_1 \mod (w_2 w_3),\quad
{\tilde n}_2 =  n_2 \mod (w_1 w_3),\quad
{\tilde n}_3 =  n_3 \mod (w_1 w_2)
\end{equation}
from which it follows that
\begin{equation}
n_1 = {\tilde n}_1 w_2 w_3 + r_1,\quad
n_2 ={\tilde n}_2 w_1 w_3 + r_2,\quad
n_3 = {\tilde n}_3 w_1w_2 + r_3
\end{equation}
where $0\le r_1 < w_2w_3$, $0 \le r_2 < w_1w_3$ and $0 \le r_3 <
w_1w_2$. This divides the energy levels into $w_1^2w_2^2w_3^2$ subsets
according to the values of $r_1, r_2$ and $r_3$. From
eq.~(\ref{eq:qenergy}), the energy levels become
\begin{equation}
E =  2 \sqrt{2k} \sum_{i=1}^3 \left[ (w_1w_2w_3) {\tilde n_i} +
     w_i(r_i + \frac{1}{2} \pm \frac{\nu_i}{2}) \right]
\end{equation}
so that the energy levels within each subset ($r_1,r_2,r_3$) have the
characteristic degeneracy of SU(3). This is illustrated by the color
coding in Fig.~\ref{fig:en2}.

\section{Summary and Conclusions}
\label{sec:con}

The {\it Caged Anisotropic Harmonic Oscillator} is a new
superintegrable Hamiltonian, namely
\begin{equation}
H = \frac{1}{2}(p_x^2 + p_y^2 + p_z^2) + k(l^2 x^2 + m^2 y^2 + n^2
z^2) + \frac{k_1}{x^2} + \frac{k_2}{y^2} + \frac{k_3}{z^2}.
\end{equation}
If the frequency multipliers are integers, then the Hamiltonian is
superintegrable. We have found the five isolating integrals for the
classical motion in three degrees of freedom. Three of the integrals
of motion -- the energies in each oscillation -- are quadratic in the
canonical momenta and arise from separation of the Hamilton-Jacobi
equation in rectangular Cartesians. The other two integrals are still
polynomial in the momenta, but now of order $2(l+m-1)$ and $2(l+n-1)$
respectively.  If $l=m=n =1$, the Hamiltonian becomes the well-studied
Smorodinsky-Winternitz system~\cite{Fr65,Ma67,Ba07,Ev90b}, and all the
integrals are then quadratic and arise from separability of the
Hamilton-Jacobi equation.

The system is interesting for at least three reasons. First, from the
perspective of integrability, there are still very few systems known
with integrals of motion that are polynomials in the momenta of higher
order than 2 \cite{Hie}. Systematic searches for Hamiltonian systems
with higher order polynomial invariants have been performed,
confirming the impression that they are rare~\cite{Thom,Eva}. Given
this sketchy and disparate information, we have no unifying theory of
the conditions for the existence of such integrals of motion

Second, from the perspective of superintegrability, if the integrals
of motion are all quadratic in the momenta, then a classification
theorem exists and all systems in flat space have been found
\cite{Fr65, Ma67, Ev90}. Such systems always arise from separability
of the Hamilton-Jacobi equation in more than one coordinate
system. However, the Caged Anisotropic Oscillator joins the Toda
Lattice and the Generalized Kepler Problem as an example of a system
for which some of the integrals are cubic polynomials or higher, and
then the superintegrability does not arise from separability in more
than one coordinate system. It would be interesting to classify such
systems and find all examples in flat space. In particular, the Caged
Anisotropic Oscillator is the second superintegrable Hamiltonian to be
deduced by the method of projection introduced in
\cite{VE08}. Essentially, the idea is to view superintegrable motion
in three degrees of freedom as a projection of a higher dimensional
superintegrable system, such as the Coulomb or Kepler problem, or the
harmonic oscillator. Are there any more such systems to be found?

Third, from the perspective of group theory in quantum mechanics, the
proper interpretation of the symmetry or degeneracy group remains
unclear. Already in 1940, Jauch \& Hill~\cite{Ja40} noted that the
quantum mechanical problem of the anisotropic oscillator presents
problems which leaves its symmetry group in doubt. Since that day,
there have been a number of different suggestions in the literature as
to the proper interpretation of the symmetry
group~\cite{Lo73,Ki73,Ro89,Bo97}. Although these procedure seem
reasonable, they are more along the lines of {\it a posteriori}
justification than compelling argument.


\end{document}